\documentclass[prl,reprint,rbibnotes,aps,showkeys,unsortedaddress,runinaddress,superscriptaddress]{revtex4-1}
\usepackage{amsmath,amssymb,epic,eepic}
\usepackage[latin1]{inputenc}
\usepackage[T1]{fontenc}
\usepackage[french,english]{babel}
\usepackage{graphicx}
\usepackage{hyperref,url}
\hypersetup{ 
    colorlinks, 
    citecolor=red, 
    filecolor=blue, 
    linkcolor=blue, 
    urlcolor=blue 
}
\usepackage{listings}
\usepackage{color}
\usepackage{textcomp}

\begin{document}
\title{Probing thermoelectric transport with cold atoms}

\author{Charles Grenier}
\affiliation{Centre de Physique Th\'eorique, Ecole Polytechnique, CNRS, 91128 Palaiseau Cedex, France.}
\author{Corinna Kollath}
\affiliation{D\'epartement de Physique Th\'eorique, Universit\'e de Gen\`eve, CH-1211 Gen\`eve, Switzerland.}
\author{Antoine Georges}
\affiliation{Centre de Physique Th\'eorique, Ecole Polytechnique, CNRS, 91128 Palaiseau Cedex, France.}
\affiliation{Coll\`ege de France, 11 place Marcelin Berthelot, 75005 Paris, France.}
\affiliation{DPMC-MaNEP, Universit\'e de Gen\`eve, CH-1211 Gen\`eve, Switzerland.}

\begin{abstract}
We propose experimental protocols to reveal thermoelectric and thermal effects in the transport 
properties of ultracold fermionic atoms, using the two-terminal setup recently realized at ETH. 
We show in particular that, for two reservoirs having equal particle numbers but different 
temperatures initially, the observation of a transient particle number imbalance during equilibration is 
a direct evidence of thermoelectric (off-diagonal) transport coefficients. 
This is a time-dependent analogue of the Seebeck effect, and a corresponding analogue of the Peltier effect can be proposed. 
We reveal that in addition to the thermoelectric coupling of the constriction a 
thermoelectric coupling also arises due to the finite dilatation coefficient of the reservoirs.
We present a theoretical analysis of the protocols, and assess their feasibility by estimating the 
corresponding temperature and particle number imbalances in realistic current experimental conditions.  
\end{abstract}


\maketitle

Thermoelectricity has been a recurrent theme in condensed matter physics. 
The increasing demand for sustainable energy sources, as well as progress in 
materials science, have triggered a marked increase of interest and research in 
the subject over the past two decades \cite{Goldsmid:1315371,macdonald2006thermoelectricity,snyder2008com,HeremansDresselhaus}.
A better understanding of fundamental processes controlling thermal and thermoelectric transport 
is likely to bring progress in the field at large.

Ultra-cold atomic gases offer a remarkably clean and controllable set-up to investigate interacting quantum systems. 
Phenomena involving the transport of atoms have been the focus of several experiments in this context, 
for example probing the reaction to external forces~\cite{4,5} as 
observing Bloch oscillations in optical lattices~\cite{6,7}, 
or the expansion dynamics of atoms in disordered~\cite{8,9,10} and lattice potentials~\cite{12}. 
The transport of impurities~\cite{PhysRevLett.85.483,PhysRevLett.103.150601,PhysRevA.85.023623} and  
spin diffusion~\cite{Sommer_Ku_Roati_Zwierlein_2011} have also been investigated. 
Recently, increasing interest has been devoted to closer analogues of mesoscopic transport devices,   
e.g. modelling `quantum pumps' and 'batteries'~\cite{17,18,19}. 
In a recent experiment transport of fermionic atoms between two 
reservoirs connected through a tunable constriction was realized~\cite{Brantut}. 

\begin{figure}[ht!]
 \centering
 \includegraphics[width=0.975\linewidth]{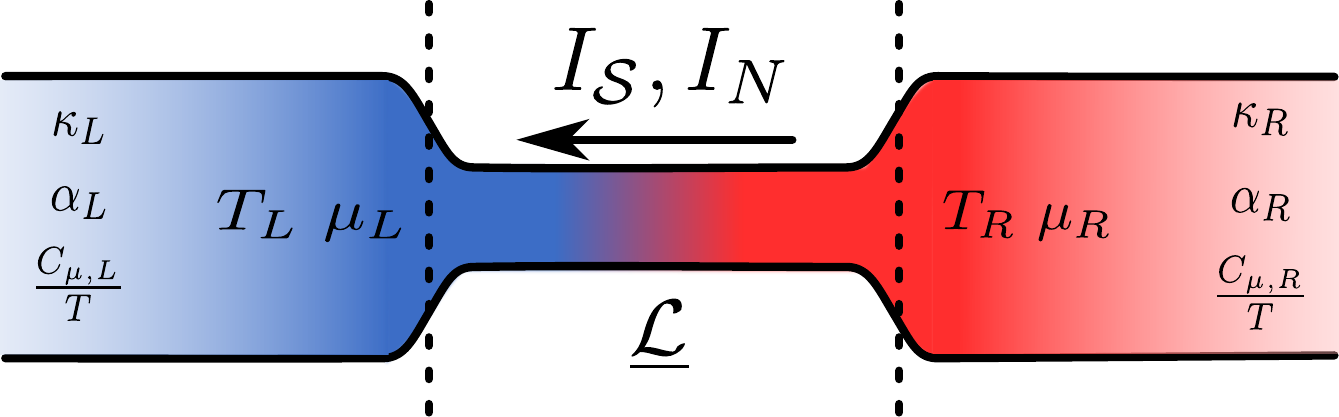}
 \caption{Two-terminal transport setup. The left (L) and right (R) reservoirs are characterized by their 
temperatures $T_{L,R}$, chemical potentials $\mu_{L,R}$ and their thermodynamic coefficients $\kappa,C_\mu,\alpha$,  
cf.~Eq.~(\ref{eq:thcoeffres_1}). 
The constriction (region between dashed lines) is characterized by its matrix of linear response coefficients $\underline{\mathcal{L}}$. 
Due to the different temperatures and chemical potentials of the reservoirs, a particle ($I_N$) and entropy ($I_{\mathcal{S}}$) current 
flow in the constriction (in our conventions, $I>0$ corresponds to a flow from right to left). 
}
\label{fig:setup}
\end{figure}

\begin{figure}
 \centering 
 \includegraphics[width=1.0\linewidth]{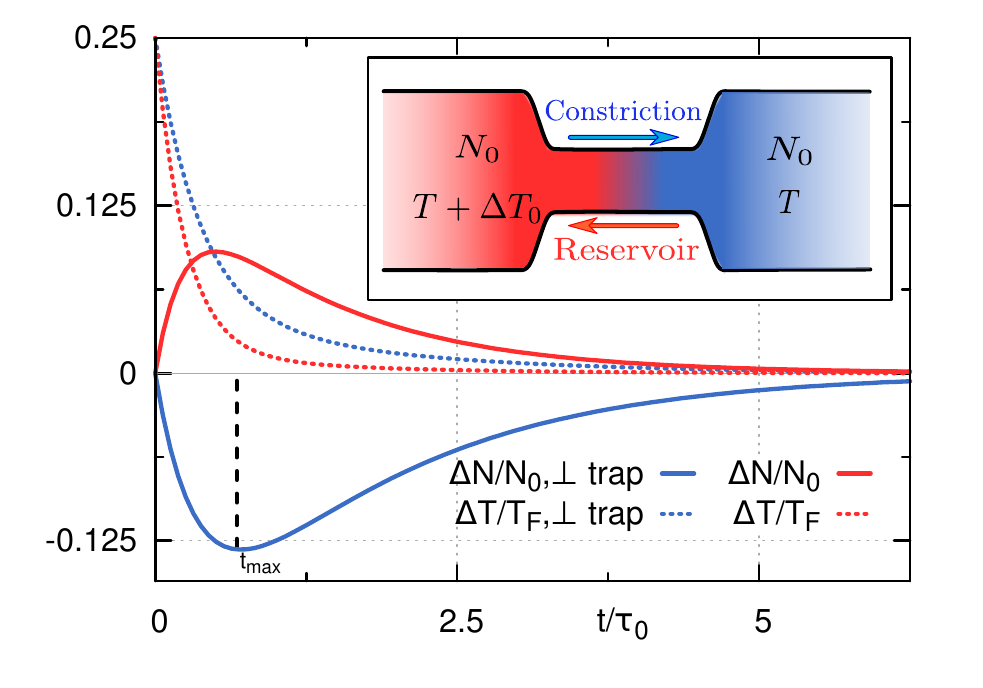}
 \caption{
{\it Inset:} 
Sketch of the proposed protocol for detecting a transient thermoelectric Seebeck-like effect.  
The two reservoirs are prepared with equal particle numbers but with an initial temperature imbalance 
$\Delta T_0$. 
A transient particle imbalance signals a non-zero Seebeck coefficient. 
The arrows indicate the direction of the particle flow at short time in the different cases (see text). 
{\it Main:}  
Particle and temperature imbalance (plain and dashed curves, respectively) as a function of time, 
 for a 2D diffusive constriction with (blue) or without (red) transverse harmonic contribution ($\perp$ trap). 
The results depicted are for $\Delta T_0/T_F=25\%$, at an initial temperature $T/T_F=0.4$.
\label{fig:Proposal}
}
\end{figure}

In this letter, we go beyond atom transport and design a proposal to observe both thermal and thermoelectric transport 
in the context of cold atoms, in the two terminal geometry of Ref.~\cite{Brantut} (Fig.~\ref{fig:setup}). 
We suggest protocols specifically geared at revealing the off-diagonal transport coefficients controlling 
thermoelectric effects (inset of Fig.~\ref{fig:Proposal}). 
These protocols can be viewed as time-dependent analogues of the Seebeck and Peltier effects. 
We estimate the magnitude of the expected effects in the simplest case of non-interacting fermions, 
for a ballistic or diffusive constriction with different geometries. 
We conclude that the proposed effects should be experimentally observable in the currently available set-up.

The experimental realization of thermal and thermoelectric transport in cold atomic gases would open many possibilities.
A particularly appealing opportunity offered by cold atomic gases is the study of these effects 
i) in the regime of high temperature where $T$ is a sizeable fraction of the Fermi temperature $T_F$ 
and 
ii) in the absence of any phonon excitations. 
Purely electronic contributions to high-temperature transport and thermopower (as captured e.g. 
by Heike's formula) have been often discussed  in the context of materials with strong electronic 
correlations~\cite{PhysRevB.13.647,heikes1961thermoelectricity} and are directly relevant to thermoelectric properties of oxide materials for example. 
In the solid state context however, separating the different contributions of electrons and phonons is highly involved. 
A setup in which phonon contributions can be suppressed is therefore invaluable to reach a deeper fundamental 
understanding of thermoelectric transport of quantum interacting particles, which may well in turn 
provide guidance for better solid-state materials design. 
%

{\it General framework}. The transport setup under consideration is depicted in Fig.~\ref{fig:setup}. 
Two reservoirs of fermionic atoms are connected by a constriction. The reservoirs are described by their temperature $T_{L,R}$ and their chemical potential  
$\mu_{L,R}$ which determine the particle number $N(T,\mu)$ and entropy $\mathcal{S}(T,\mu)$ of each reservoir through the 
grand-canonical equation of state. 
Important thermodynamic properties of the reservoirs are the compressibility $\kappa$, the dilatation 
coefficient $\alpha$ and the heat capacity $C_{\mu,N}$ (at constant $\mu$ or $N$), defined by:  
\begin{equation}
\label{eq:thcoeffres_1}
 \kappa = \left. \frac{\partial N}{\partial \mu}\right|_T ,\,
 \alpha = \left.\frac{\partial N}{\partial T}\right|_\mu=\left.\frac{\partial\mathcal{S}}{\partial\mu}\right|_T ,\,
\frac{C_{\mu,N}}{T} =\left. \frac{\partial\mathcal{S}}{\partial T}\right|_{\mu,N}
\end{equation}
The particle and entropy currents flowing 
through the constriction $I_{N}=\frac{d}{dt} (N_L-N_R),\, I_{\mathcal{S}} = 
\frac{d}{dt} (\mathcal{S}_L-\mathcal{S}_R)$\footnote{The definition of the currents differs by a factor two compared
to \cite{Brantut}.} are related to 
the (small) chemical potential and temperature differences $\Delta \mu=\mu_L-\mu_R$ and $\Delta T=T_L-T_R$ by:
\begin{equation}
 \label{eq:entropy_particle_current_definition}
 \left(
\begin{array}{c}
 I_N\\
 I_{\mathcal{S}}
\end{array}
\right)
=
\underline{\mathcal{L}}
 \left(
\begin{array}{c}
\Delta \mu\\
\Delta T
\end{array}
\right) , 
\textrm{with}\quad\underline{\mathcal{L}} = 
 \left(
\begin{array}{cc}
 \mathcal{L}_{11} & \mathcal{L}_{12}\\
 \mathcal{L}_{12} & \mathcal{L}_{22}
\end{array}
\right)\,
\end{equation}
$\underline{\mathcal{L}}$ is the matrix of transport coefficients associated with the constriction. Its symmetry is 
insured by Onsager's relations~\cite{Callen:Thermodynamics,PhysRev.38.2265,PhysRev.37.405}. 

We consider a time-dependent process in which the reservoirs are prepared with given initial particle numbers and temperatures, 
and  equilibrate through exchange of particles and entropy through the constriction. 

Using (\ref{eq:entropy_particle_current_definition}) and the properties of the reservoirs given by 
(\ref{eq:thcoeffres_1}), we derive equations ruling the time evolution of the particle and temperature imbalance:  
\begin{equation}
 \tau_0\frac{d}{dt}
\left(
 \begin{array}{c}
 \Delta N/\kappa\\
  \Delta T 
 \end{array}
\right)
=
-\underline{\Lambda}
\left(
 \begin{array}{c}
 \Delta N/\kappa\\
  \Delta T 
 \end{array}
\right) , 
\underline{\Lambda}=\begin{pmatrix}
1 & -S\\ 
-\frac{S}{\ell} & \frac{L+S^2}{\ell}
\end{pmatrix}\,.
\label{eq:Transport_equation}
\end{equation}
These equations resemble the discharge of a capacitor (the reservoirs) in a resistor (the constriction), 
taking into account thermal transport as well. 
The characteristic time-scale $\tau_0\equiv\kappa/\mathcal{L}_{11}$ corresponds to the time-scale 
$R\mathcal{C}$ in a capacitor, with $R=1/\mathcal{L}_{11}$ the resistance of the constriction and $\mathcal{C}$ 
($\sim\kappa$) the capacitance~\cite{Brantut}.  
The Lorenz number~\cite{Ashcroft} of the constriction 
$L\equiv\mathcal{L}_{22}/\mathcal{L}_{11}-\left(\mathcal{L}_{12}/\mathcal{L}_{11}\right)^2$ 
is the ratio $R/(T R_T)$ of the resistance to thermal resistance.   
The thermodynamic coefficient 
$\ell\equiv C_\mu/\kappa T-\left(\alpha/\kappa\right)^2=C_N/\kappa T$ 
characterizes the reservoir and has a form similar to $L$. 
Finally, $S\equiv\alpha/\kappa-\mathcal{L}_{12}/\mathcal{L}_{11}$  
is the effective thermoelectric (Seebeck) coefficient. 
Note that $S$ has the dimension of the Boltzmann constant $k_B$, while $L$ and $\ell$ have dimension 
$k_B^2$ and $\kappa$ the dimension of the inverse of an energy.  

Two remarks are in order. 
(i) In the absence of any thermoelectric effect ($S=0$), the time constants for particle and thermal relaxation
are $\tau_0$ and $\tau_0 L/\ell$, respectively. 
At low temperature, the usual form of the Wiedemann-Franz law, when applicable, dictates that 
$L\rightarrow\pi^2/3$ and the thermodynamics of a free Fermi gas yields the same limit for  
$\ell\rightarrow\pi^2/3$. 
Hence, in this case, the Wiedemann-Franz law expresses the fact that the timescales for particle and heat relaxation are 
identical when $S=0$, an interpretation which to our knowledge has not been formulated before. 

(ii) The coupling between heat and particle transport (off-diagonal elements of $\underline{\Lambda}$) is determined by the effective Seebeck coefficient $S$. 
In the present context, this coefficient has two competing contributions: one from 
the constriction $S_c=-\mathcal{L}_{12}/\mathcal{L}_{11}$
and one from the reservoirs $S_r=\alpha/\kappa$. 
In particular, the presence of $S_r$ induces a coupling between thermal and electric transport even when the off-diagonal 
transport coefficient $\mathcal{L}_{12}$ of the constriction can be neglected. 

\begin{figure}
\centering 
\includegraphics[width=1.0\linewidth]{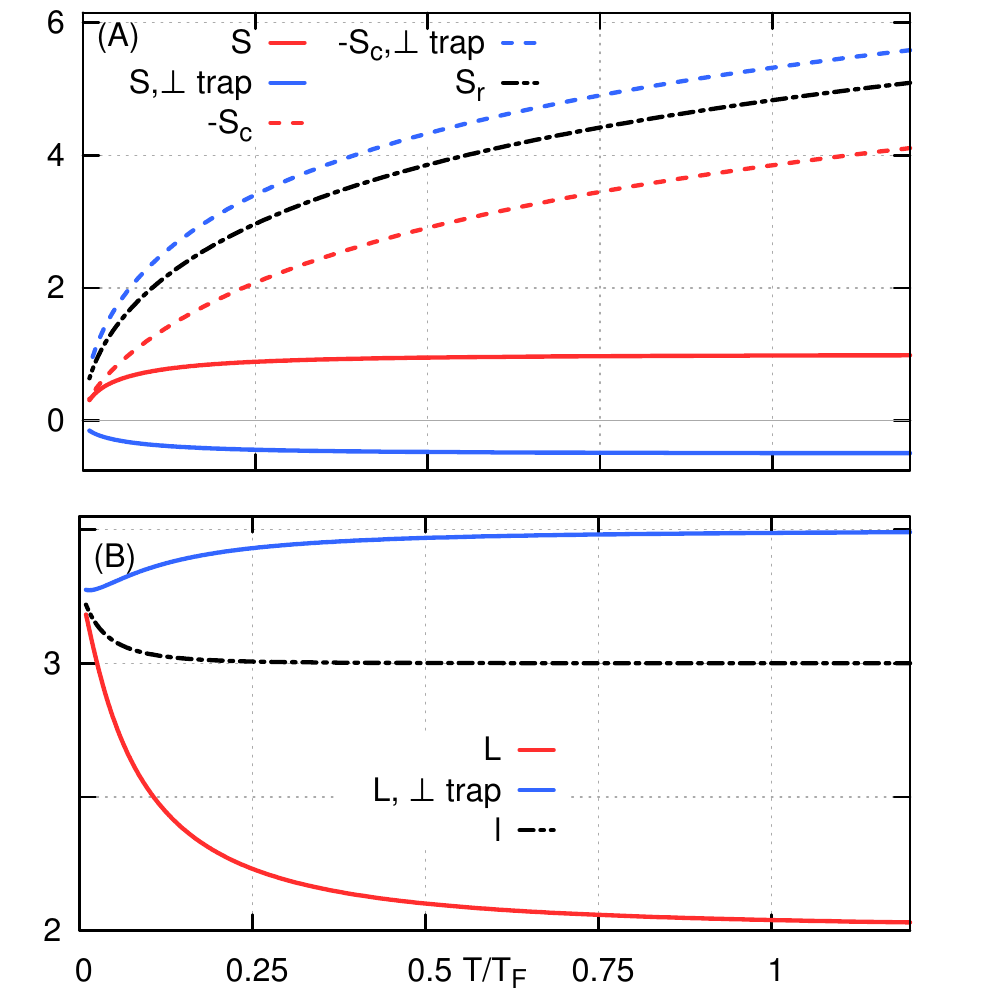}
\caption{Transport coefficients as a function of $T/T_F$ for a 2D diffusive constriction, with (blue) 
or without (red) transverse confinement, and harmonic trapping in the reservoirs.\\ 
(A)- Contributions from the reservoirs ($S_r$, in units of $k_B$) and from the constriction ($S_c$, in units of $k_B$) to the total 
effective Seebeck coefficient $S=S_r+S_c$. 
(B)- Lorenz number $L$ (in units of $k_B^2$) of the constriction, and thermodynamic coefficient $\ell$ (in units of $k_B^2$) of the trapped reservoirs.}
\label{fig:Transport_qties}
\end{figure}

On Fig~\ref{fig:Transport_qties}, we display $S_c$, $S_r$, $L$ and $\ell$ as a function of $T/T_F$, 
calculated within a Landauer-B\"{u}ttiker~\cite{RevModPhys.71.S306,Buttiker:1988:SEC:49387.49388,citeulike:213956}
formalism for a diffusive constriction (see below for details). 
The plot illustrates the competition between the contribution of the reservoir $S_r$ and that of the 
constriction $S_c$. In the case of Fig.~\ref{fig:Transport_qties}, 
this results in a positive (negative) total Seebeck coefficient in the absence (presence) of transverse harmonic 
confinement in the constriction. 
Note also on Fig.~\ref{fig:Transport_qties} the deviations from the Wiedemann-Franz law at finite temperature. 

The general solution of Eq.~\eqref{eq:Transport_equation} reads, given an initial 
particle and temperature difference $\Delta N_0$ and $\Delta T_0$:
\begin{widetext}
\begin{eqnarray}
\label{eq:particle_number_result}
  \Delta N(t) &= \left\lbrace \frac{1}{2}\left[e^{-t/\tau_-}+e^{-t/\tau_+}\right]
-\left[1-\frac{L+S^2}{\ell}\right]
\frac{e^{-t/\tau_-}-e^{-t/\tau_+}}{2(\lambda_+-\lambda_-)}
\right\rbrace \Delta N_0
+\frac{S\kappa}{\lambda_+-\lambda_-}\left[e^{-t/\tau_-}-e^{-t/\tau_+}\right]\Delta T_0\\
\label{eq:temperature_result}
  \Delta T(t) &= \left\lbrace 
\frac{1}{2}\left[e^{-t/\tau_-}+e^{-t/\tau_+}\right] 
-\left[\frac{L+S^2}{\ell}-1\right]\frac{e^{-t/\tau_-}-e^{-t/\tau_+}}{2(\lambda_+-\lambda_-)}\right\rbrace \Delta
T_0+\frac{S}{\ell\kappa(\lambda_+-\lambda_-)}\left[e^{-t/\tau_-}-e^{-t/\tau_+}\right]\Delta N_0
\end{eqnarray}
\end{widetext}
The inverse time-scales $\tau_\pm^{-1}=\tau_0^{-1}\lambda_\pm$ are given by the eigenvalues of the transport matrix $\underline{\Lambda}$
\begin{equation}
 \label{eq:timescales}
 \lambda_\pm = \frac{1}{2}\left(1+\frac{L+S^2}{\ell}\right)\pm\sqrt{\frac{S^2}{\ell}+\left(\frac{1}{2}-\frac{L+S^2}{2\ell}\right)^2}\,.
\end{equation} 

In principle, eqs. \eqref{eq:particle_number_result} and \eqref{eq:temperature_result} enable one to extract the 
thermodynamic and transport coefficients from experimental measurements. In particular, the Wiedemann-Franz law and its possible violations
could be tested.
 However, thermoelectric effects are more difficult to extract. This is due to the fact
that in the presence of both a particle and temperature imbalance at $t=0$, the time-evolution is typically dominated by 
the exponential decay involving the symmetric combination of exponentials, 
and terms responsible for offdiagonal transport are masked.

Hence, we propose two specific protocols in order to reveal thermoelectric effects, estimate the 
expected magnitude of the signal and confirm experimental feasibility.    


{\it Experimental proposal for off-diagonal (thermoelectric) transport}. 
The first protocol focuses on the particle current induced by a temperature difference, and is a transient analogue of the Seebeck effect.   
The system is prepared with equal number of particles in the two reservoirs $\Delta N_0=0$, but with a temperature difference (see inset of Fig.~\ref{fig:Proposal}). The temperature difference can for example be prepared by heating one of the reservoirs 
({\it eg} by using laser light) with a closed constriction, which prevents particle transfer.
After reopening the constriction, the off-diagonal coupling between particle and heat transport will lead to an 
evolution of the particle number difference given by : 
\begin{equation}
  \frac{\Delta N(t)}{N_0}=\frac{S\,T_F\,\kappa}{N_0}\,\frac{e^{-t/\tau_-}-e^{-t/\tau_+}}{\lambda_+-\lambda_-}\frac{\Delta T_0}{T_F}\,,
\label{eq:particle_number_SG_result}
\end{equation}
where $N_0,\, T_F$ refer to reservoirs at equilibrium.
Hence, a transient particle imbalance during equilibration is a `smoking gun' 
observation revealing the existence of a non-zero Seebeck coefficient.
Since typically $\tau_+<\tau_-$, the particle imbalance first reaches an extremum at a time $t_{max} = \tau_0 (\ln\lambda_{+}-\ln\lambda_{-})/(\lambda_+-\lambda_{-})$
and then at long times relaxes exponentially to zero with a characteristic time $\tau_-$. 
This behaviour is exemplified in Fig.~\ref{fig:Proposal}. The sign of the particle imbalance is given by the 
sign of the effective Seebeck coefficient and thus depends on the considered situation. 
If the reservoir effect is the dominant mechanism ($S>0$), then particles tend at first to flow 
from the cold to the hot reservoir 
(see Fig.~\ref{fig:Proposal}), due to its lower chemical potential. 
This behaviour is in contrast to the classical intuition of 
particles flowing from the high pressure (warmer) to the low pressure (colder) side.  
On the contrary, if the thermoelectric properties of the constriction dominate ($S<0$), then particles first flow from 
the hot to the cold reservoir. 
In both cases the temperature imbalance equilibrates monotonically. As expected, the entropy current 
($I_{\mathcal{S}} = SI_N-\Delta T/R_T T$) is flowing from the hot to the cold reservoir, 
because the contribution of thermal diffusion to entropy flow (second term) always dominates 
over the thermoelectric contribution from particle transport (first term).    
A second complementary protocol can be proposed in which the system is prepared initially with a particle number 
imbalance but equal temperatures in the two reservoirs. 
This is the analogue of a Peltier experiment, in which temperature and particle number 
imbalances have reversed roles.

In order to estimate the magnitude of the expected effects and assess whether they can be observed 
experimentally, we have considered the simplest case of non-interacting fermions and computed the 
transport coefficients of the constriction using a Landauer-B\"{u}ttiker formalism, in the 
spirit of mesoscopic physics~\cite{RevModPhys.71.S306,Buttiker:1988:SEC:49387.49388,citeulike:213956}. 
In this framework, the transport coefficients $\mathcal{L}_{ij}$ of the constriction can
be expressed in terms of the moments $M_n$ of an energy-dependent (dimensionless) transport function $\Phi(\epsilon)$ as: 
$\mathcal{L}_{11}=2M_0/h$, $\mathcal{L}_{22} = 2k_B^2/h\cdot M_2/(k_BT)^2$ and 
$\mathcal{L}_{12}= 2k_B/h\cdot M_1/(k_BT)$, with:
\begin{equation}
 \label{eq:transport_coeffs}
 M_n = \int_0^\infty d\epsilon\, \Phi(\epsilon)\left(-\frac{\partial f}{\partial \epsilon}\right)(\epsilon-\mu)^n 
\end{equation}
where $f$ is the equilibrium Fermi function. 
The precise form of the transport function $\Phi(\epsilon)$ depends on the dispersion relation and on 
the scattering mechanism in the constriction. We used a relaxation time approximation with an energy-independent 
scattering time, and considered either a ballistic or a diffusive constriction (both cases can be realized 
experimentally~\cite{Brantut} using speckle noise of adjustable strength).
The constriction is two-dimensional, with or without harmonic confinement in the transverse direction, and 
the reservoirs are taken as three dimensionally harmonically trapped non-interacting Fermi gases.
%

\begin{figure}
 \centering 
 \includegraphics[width=1.0\linewidth]{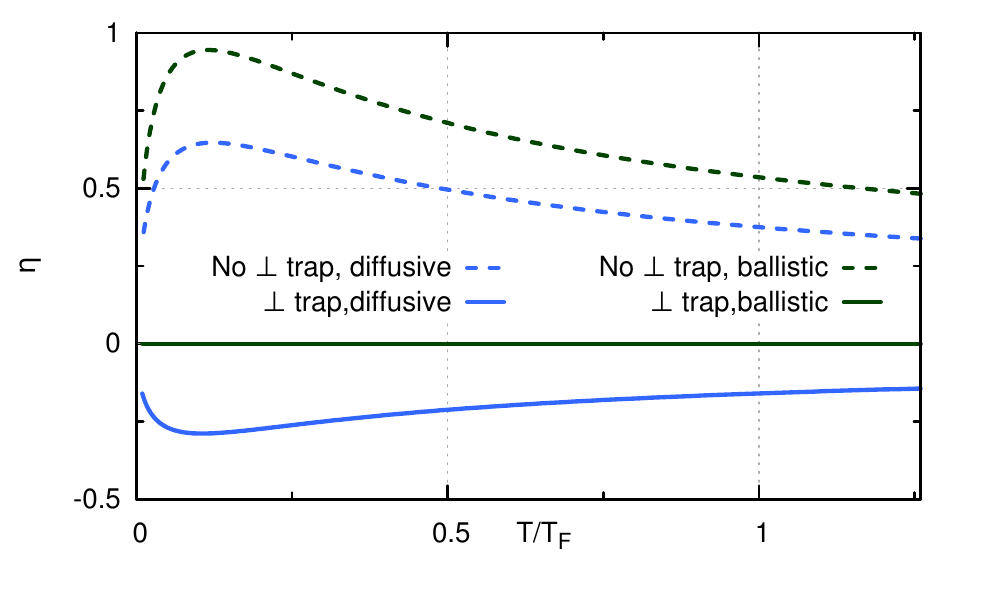}
 \caption{Efficiency $\eta$ (see text) as a function of $T/T_F$ for a ballistic or diffusive 2D constriction, 
 with or without transverse confinement (with harmonic trapping of the reservoirs).}
 \label{fig:eta_vs_T}
\end{figure}

%
In order to observe the particle imbalance which is induced by the temperature imbalance, its amplitude should be sizable. 
For the typical parameters used in Fig.~\ref{fig:Proposal}, 
The amplitude of the maximum $\Delta N(t_{max})/N_0$ is seen to exceed $10\%$ , which makes it within reach of current experimental setups~\cite{Brantut}.
To further quantify the effects, we define a parameter $\eta\equiv\left(\Delta N (t_{max})/N_0\right)\cdot\left(T_F/\Delta T_0\right)$ 
which can be understood as a thermoelectric efficiency of the system. It compares the maximum relative particle imbalance 
to the initial temperature imbalance in units of the Fermi temperature. 
A similar quantity can be defined for the Peltier-like protocol, leading to an efficiency $\eta_P=\eta/l$ 
(hence expected to be smaller than $\eta$ since $l>1$ - see Fig.~\ref{fig:Transport_qties}). 
Figure~\ref{fig:eta_vs_T} displays $\eta$ as a function of $T/T_F$, comparing four cases - ballistic and diffusive and 
with or without transverse confinement in the constriction. 
In particular, one sees that $\eta \simeq 50 \%$ at a commonly reached experimental temperature $T/T_F=0.4$~\cite{Brantut}, 
meaning that an initial relative temperature imbalance $\Delta T_0/T_F=20\%$ with reservoirs containing $N_0 = 100000$ particles would lead to a 
maximal particle imbalance of the order of $10000$ particles, which is experimentally sizeable. 
We note (Fig~\ref{fig:eta_vs_T}) that a very peculiar situation arises for a two-dimensional ballistic constriction with 
harmonic confinement. In this case, $S_r$ and $S_c$ perfectly compensate each other, leading to $S=\eta=0$. 
This is due to the fact that, in that case (assuming an energy-independent collision time), 
the transport function $\Phi(\epsilon)$ has exactly the same energy dependence than that of the 
density of states in the reservoirs ($\sim\epsilon^{2}$). 
While this perfect cancellation relies on specific assumptions of our simplified modelling, the qualitative 
conclusion is expected to be robust: in this particular case the thermoelectric effects are expected to be 
quite small. 
%

To summarize, we have shown that thermoelectric effects can be measured in cold atomic gases within the setup of Ref.~\cite{Brantut}.
In the current experimental temperature regime, offdiagonal transport properties arise from a combination of the
thermoelectric properties of the constriction and the finite dilatation coefficient of the Fermi gas in the reservoirs.
Fundamental questions about high-temperature transport can be addressed in this framework. 

Inter-particle interactions have been neglected here, implicitly assuming that they have been turned off using a 
Feschbach resonance. A natural extension of this work is thus to include interactions,
and to go beyond the simple Drude description of transport in the constriction. 
One can also think of reintroducing phonons in a controlled manner by 
simulating their action via a bosonic bath, 
an important ingredient for establishing contact with thermoelectric properties of solid-state materials.
Other possible directions include the effect of a lattice in the constriction, or a modification of the geometry of the constriction, 
which  would provide a fruitful relation to recent developments on thermoelectric devices in the mesoscopic physics context~
\cite{PhysRevLett.102.146602,PhysRevB.85.075412}.
 
{\it Note added} : 
As the writing of manuscript was being completed, we became aware of a recent preprint by H. Kim and D. Huse~\cite{Huse2012thermoelectricity} 
considering spin and heat transport in a cold Fermi gas.  
\acknowledgements
We thank J.-P.~Brantut, S.~Krinner, J.~Meineke, D.~Stadler and T.~Esslinger for invaluable discussions. 
Support was provided by the ANR (project FAMOUS), the Triangle de la Physique (project CORSA),
the DARPA-OLE program, the SNSF (Division II, MaNEP) and the NSF (under grant NSF-PHY11-25915). 
The hospitality of ETH and KITP are gratefully acknowledged. 

\bibliography{Paper_th} 
\end{document}